\documentclass[11pt,a4paper]{article}
\usepackage[T1]{fontenc}
\usepackage[utf8]{inputenc}
\usepackage{amssymb,amsbsy,amsmath,amsthm,amsfonts}     % math stuff
\usepackage{stmaryrd}                                   % additional math symbols
\usepackage{graphicx}
\usepackage[textfont=footnotesize,labelfont={footnotesize,bf}]{caption}
\usepackage[position=top,caption=false]{subfig}
\usepackage{booktabs}
\usepackage{enumitem}
\usepackage[dvipsnames]{xcolor}
\usepackage[top=30mm,right=30mm,left=30mm,bottom=30mm]{geometry}
\usepackage[colorlinks=true,linkcolor=blue,urlcolor=blue]{hyperref}
\usepackage[affil-it,auth-sc]{authblk}
\usepackage[colorinlistoftodos,textwidth=25mm,textsize=footnotesize]{todonotes}
\usepackage[normalem]{ulem}                             % strikeout text
\usepackage{lipsum}                                     % generate dummy text
\usepackage[
    backend=bibtex,
    style=numeric,
    citestyle=numeric-comp,
    maxbibnames=99,
    minbibnames=99,
    maxcitenames=2,
    mincitenames=1,
    giveninits=true,
    sorting=none,
    sortlocale=auto,
    natbib=true,
    url=false, 
    isbn=false,
    doi=true,
    eprint=true
]{biblatex}
\hypersetup{colorlinks=true,linkcolor=blue,urlcolor=blue}
\newcommand{\beq}{\begin{equation}}
\newcommand{\eeq}{\end{equation}}

\newcommand{\beqar}{\begin{eqnarray}}
\newcommand{\eeqar}{\end{eqnarray}}
\newcommand{\bit}{\begin{itemize}}
\newcommand{\eit}{\end{itemize}}
\newcommand{\benum}{\begin{enumerate}}
\newcommand{\eenum}{\end{enumerate}}
\newcommand{\barr}{\begin{array}}
\newcommand{\earr}{\end{array}}

\def\scalp{\boldsymbol{\,\cdot\,}}

\def\XXint#1#2#3{{\setbox0=\hbox{$#1{#2#3}{\int}$}
   \vcenter{\hbox{$#2#3$}}\kern-.5\wd0}}

\def\b0{\boldsymbol{0}}

\def\bc{\boldsymbol{c}}

\def\be{\boldsymbol{e}}

\def\bk{\boldsymbol{k}}

\def\bv{\boldsymbol{v}}

\def\bA{\boldsymbol{A}}

\def\bK{\boldsymbol{K}}

\def\f0{\ensuremath{\mathbb{O}}}

\graphicspath{{./}{./figures/}}                         % path to figures folder

\title{Prestress tuning of negative refraction and\\ wave channeling from flexural sources}

\author[1]{G. Bordiga}
\author[1]{L. Cabras}
\author[1]{A. Piccolroaz}
\author[1]{D. Bigoni\footnote{Corresponding author: e-mail: \href{mailto:bigoni@ing.unitn.it}{bigoni@ing.unitn.it}; phone: +39\,0461\,282507.}}

\affil[1]{Department of Civil, Environmental and Mechanical Engineering, University of Trento, Italy}

\date{}

\begin{document}

\maketitle

\begin{abstract}
\noindent
The quest for wave channeling and manipulation has driven a strong research effort on topological and architected materials, capable of propagating localized electromagnetical or mechanical signals.
With reference to an elastic structural grid, which elements can sustain both axial and flexural deformations, it is shown that material interfaces can be created with structural properties tuned by prestress states to achieve total reflection, negative refraction, and strongly localized signal channeling.
The achievement of a flat lens and topologically localized modes is demonstrated and tunability of the system allows these properties to hold for a broad range of wavelengths. 
An ingredient to obtain these effects is the use, suggested here and never attempted before, of concentrated pulsating moments.
The important aspect of the proposed method is that states of prestress can be easily removed or changed to tune with continuity the propagational characteristics of the medium, so that a new use of vibration channeling and manipulation is envisaged for elastic materials. 
\end{abstract}


\section{Introduction}
\label{sec:introduction}
The possibility of channeling, trapping, and controlling waves opens new possibilities such as cloaking of a part of a body, or achieving total reflection and negative refraction, effects which prelude the realization of flat lens, able to overcome the diffraction limits through superlensing effects~\cite{veselago_1968,pendry_2000,kaina_2015,craster_2012}.
Negative refraction in elastic lattices has been obtained for a plane wave~\cite{zhang_2004,brun_2010,piccolroaz_2017,piccolroaz_2017a,cabras_2017} (and experimentally confirmed~\cite{morvan_2010,zhu_2014,liu_2017}), while only in electromagnetism~\cite{lin_2017}  and in elastic plates under flexure~\cite{farhat_2010} pulsating sources have been considered so far. 

Edge waves and trapped modes have been demonstrated, but their realization involves the use of gyroscopic systems~\cite{wang_2015,carta_2017,garau_2018}, piezoelectric elements~\cite{celli_2015}, topological materials~\cite{xia_2018} or structures inducing floppy modes~\cite{mao_2018,ma_2018}. 
Recently topologically protected edge waves for plates subject to flexure have been demonstrated~\cite{miniaci_2018}.

A route to achieve wide frequency bandwidth for the above-mentioned dynamical effects is tunability of the mechanical properties, representing a crucial ingredient in the development of metamaterials or architected materials, so that the wave propagation can be changed and manipulated, according to different needs, for instance in a way that an interface may be occasionally made permeable to mechanical disturbances or changed to realize total reflection or otherwise to allow negative refraction. 
Tunability has been addressed with reconfigurable origami materials~\cite{overvelde_2017}, connectivity~\cite{wang_2015}, piezoelectric effects~\cite{celli_2015}, or, finally, prestress~\cite{bigoni_2008,gei_2009,pal_2018}. 
This latter technique can simply be implemented by applying forces to a structure prior to wave propagation, which results strongly influenced, as the dynamics of musical instruments clearly show. 
Forces can be readily applied and removed, so that tunability can be easily and quickly obtained. 

Prestress is shown in this article to govern flexural and axial waves propagation in an elastic square grid of beams, so that axial forces can be applied to a set of beams arranged in a layer inside an infinite lattice of beams not subject to prestress. 
For a certain level of prestress (always assumed tensile to avoid buckling), the layer is shown to completely reflect waves, while waves are transmitted for a different prestress level and may display negative refraction and focusing. 
Moreover, narrow layers of prestressed elements give rise to highly localized trapped modes, showing strongly focused propagation.

\section{Time-harmonic vibration of a lattice of axially and flexurally deformable beams}
\label{sec:mathematical_setting}
\begin{figure}[htb]
\centering
\subfloat{
    \label{fig:lattice}
    \includegraphics[width=0.6\linewidth]{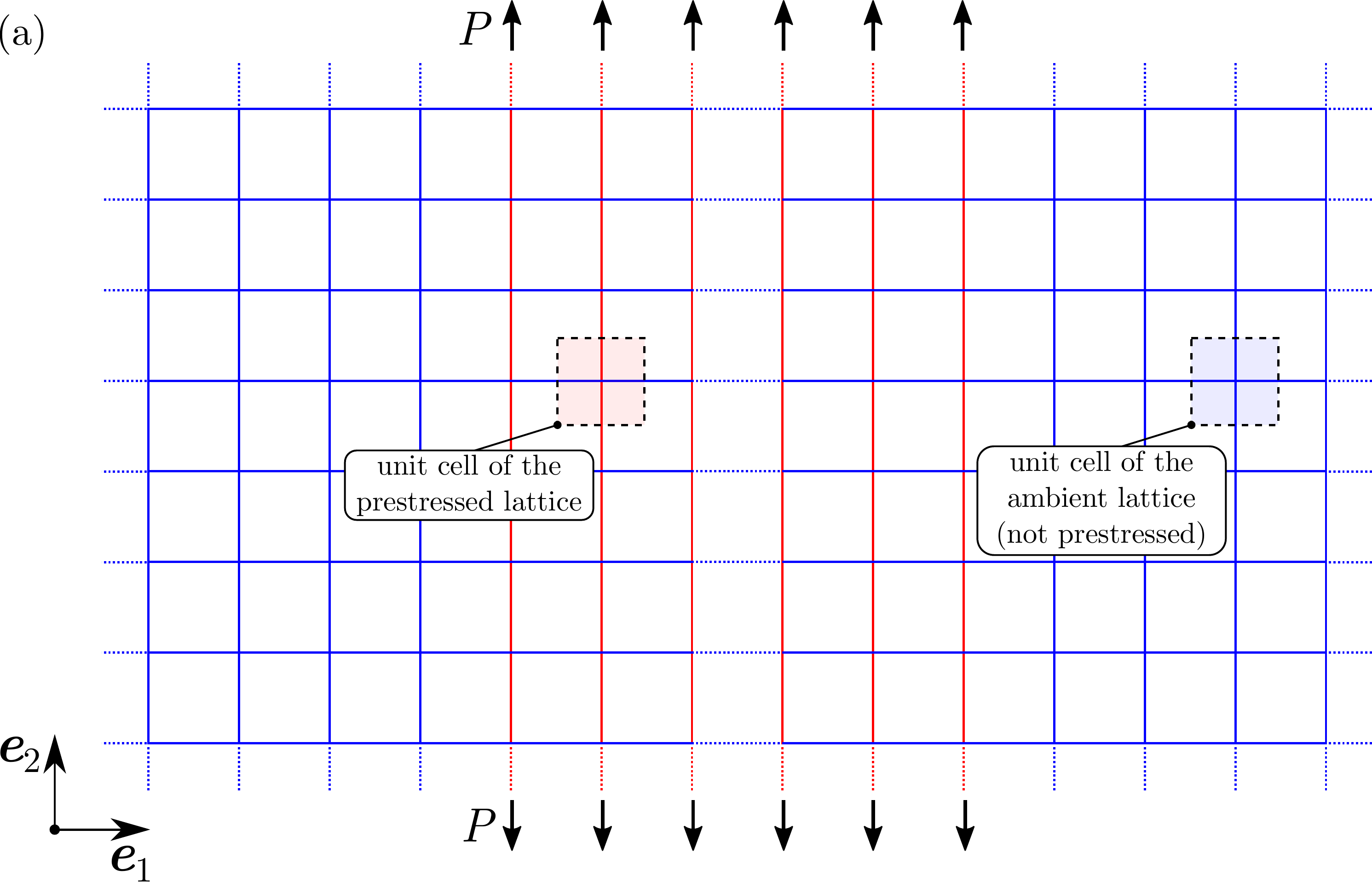}
}\\
\subfloat[][$p=0$]{
    \label{fig:dispersion_surface_p_0}
    \includegraphics[width=0.22\linewidth]{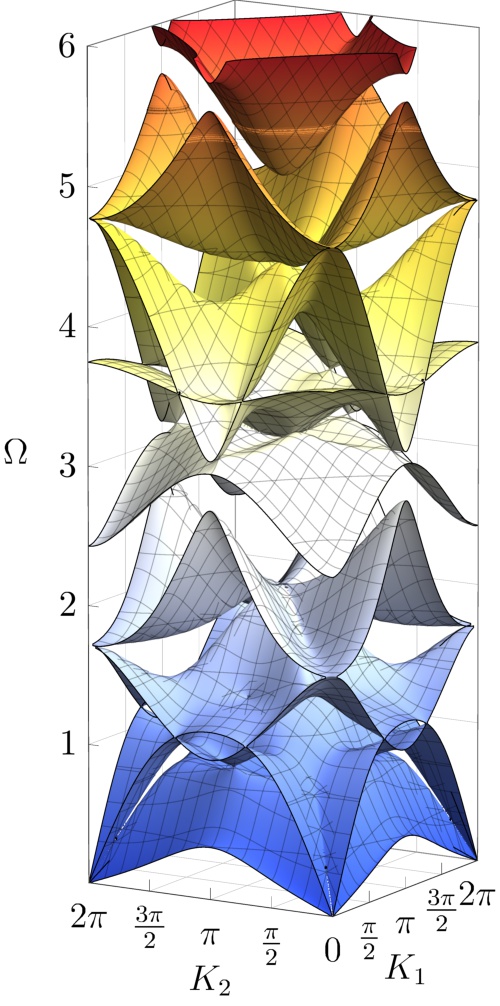}
}
\subfloat[][$p=65$]{
    \label{fig:dispersion_surface_p_65}
    \includegraphics[width=0.22\linewidth]{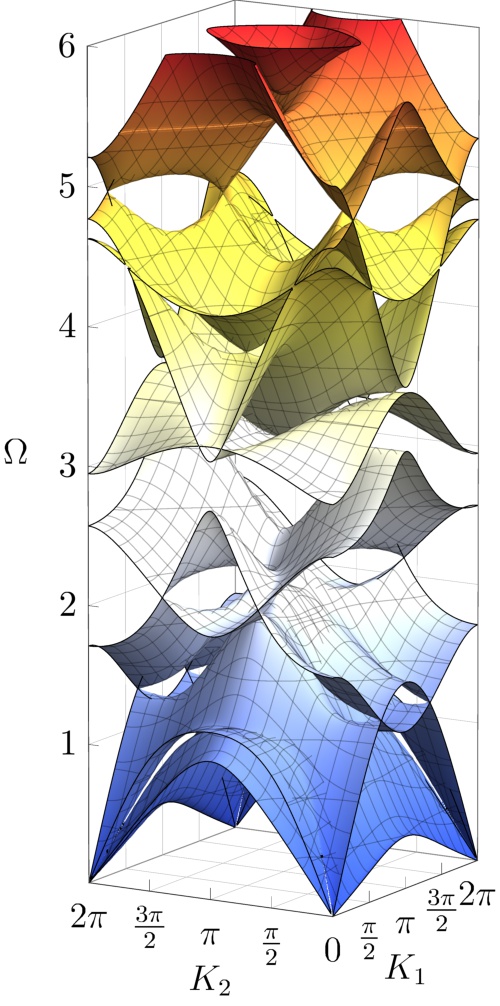}
}
\subfloat[][$p=190$]{
    \label{fig:dispersion_surface_p_190}
    \includegraphics[width=0.22\linewidth]{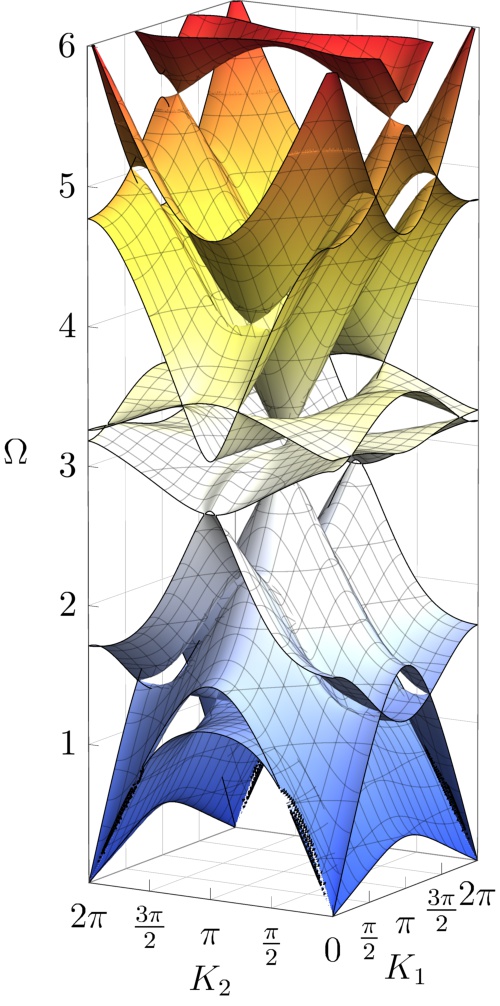}
}
\caption[]{\label{fig:lattice_dispersion_surfaces}\subref{fig:lattice} Geometry of the beam grid lattice and the prestressed layer. 
The vertical beams marked in red are subject to the axial prestress $P$, so that a tunable interface is realized. 
The dispersion surfaces of the ambient lattice (not prestressed) and of the prestressed lattice are reported in parts~\subref{fig:dispersion_surface_p_0},~\subref{fig:dispersion_surface_p_65}, and~\subref{fig:dispersion_surface_p_190}, showing the strong effect of the prestress (made dimensionless as $p=PL^2/EI$).
}
\end{figure}

An infinite square lattice of elastic Rayleigh beams~\cite{piccolroaz_2014} (both axially and flexurally deformable and of length $L$) is assumed, where layers are subject to a prestress, induced by axial forces $P$, acting on a set of rods (Fig.~\ref{fig:lattice}). 
The prestressed layer defines a `structured interface', with a width chosen as $40 L$. 
% 
%%%%%%%%%%%%%%%%%%%%%%%%%%%%%%%%%%%%%%%%%%%%%%%%%%%%%%%%%%%%%%%%%%%%%%
%%%%%%%%%%%%%%%%%%%%%%%%%%%%%%%%%%%%%%%%%%%%%%%%%%%%%%%%%%%%%%%%%%%%%%
% 
The slenderness $\lambda=L/r$ (where $r$ is the radius of gyration of the beam's cross-section) of each beam is assumed equal to $15$ and the lattice is analyzed for time-harmonic vibration, with angular frequency $\omega$, so that introducing a dimensionless local coordinate $\xi=s/L$, the equations governing the dynamics of the lattice in terms of axial and transverse displacements, $u$ and $v$, are
\begin{align}
\label{eq:governing_equations_u}
    u''(\xi) + \Omega^2 \, u(\xi) &= 0 \,, \\
\label{eq:governing_equations_v}
    v''''(\xi) + (\Omega^2-p) \, v''(\xi) - \lambda^2\,\Omega^2 \, v(\xi) &= 0 \,,
\end{align}
where $\Omega=\omega\,L\sqrt{\rho/E}$ and $p=PL^2/EI$ are the dimensionless angular frequency and axial prestress, respectively ($\rho$ is the mass density, $E$ the Young modulus, $I$ the second moment of the cross section's area and the prime denotes differentiation with respect to $\xi$). 

An exact Floquet-Bloch analysis guides the determination of the level of prestress to tune desired dynamic responses of the lattice, for instance, achieving the total reflection of a wave, or its negative refraction, when the wave impinges on an interface separating the lattice without prestress from the prestressed layer.
In particular, the prestress level is determined using the slowness contours, obtained from the dispersion equation, of the periodic homogeneous lattice without and with prestress (dispersion surfaces as influenced by the prestress $p$ are shown in Fig. \ref{fig:dispersion_surface_p_0}--\subref*{fig:dispersion_surface_p_190}).
Special attention has been paid to eliminate the possibility of buckling, by selecting a \textit{tensile} prestress (even though effects similar to those shown in the following can be obtained for compressive prestress, or changing the slenderness of a layer of beams, an option presented in the supplementary material).

The eigenvalue problem governing wave propagation in the periodic grid can be easily formulated as follows (see also~\cite{bordiga_2018} for details).
On each beam of the unit cell, the solution of~eqs.~\eqref{eq:governing_equations_u}~and~\eqref{eq:governing_equations_v} can be expressed in terms of a linear combination of complex exponentials, $u(\xi) = C\,e^{i\,\eta\,\xi}$ and $v(\xi) = D\,e^{i\,\gamma\,\xi}$, where the characteristic roots are
\begin{align*}
    \eta_{1,2} &= \pm \Omega \,, \\ 
    \gamma_{1,2,3,4} &= \pm\sqrt{\frac{1}{2}\left(\Omega^2-p \pm \sqrt{4\lambda^2\Omega^2+(\Omega^2-p)^2}\right)} \,.
\end{align*}
Then, the solution of the unit cell is constrained by imposing the junction and equilibrium conditions at the central joint and the Bloch-Floquet boundary conditions between corresponding sides of the unit cell for: (i.) axial and flexural displacements, (ii.) rotation, (iii.) internal moment, (iv.) axial and shear forces.  

A homogeneous linear system of equations governing the propagation of Floquet-Bloch waves is found in the form 
\begin{equation}
\label{eq:system}
    \bA(\Omega, \bK, p, \lambda) \, \bc = \b0 \,,
\end{equation}
where $\bA(\Omega, \bK, p, \lambda)$ is a $24\times24$ complex matrix, function of the angular frequency $\Omega$ and the wave vector $\bK$, of dimensionless components $K_1$ and $K_2$ (obtained from multiplication of the Bloch wave vector, $\bk = k_1\be_1 + k_2\be_2$, by $L$), as well as the prestress parameter $p$ and the slenderness $\lambda$.
Finally, vector $\bc$ defines the waveform as it collects the 24 complex constants that multiply the exponential functions appearing in the displacement fields.
As the system~\eqref{eq:system} is homogeneous, all non-trivial solutions are found when the matrix $\bA(\Omega,\bK)$ becomes singular, a condition providing the dispersion equation.
The latter equation has been solved numerically in order to identify the influence of the axial prestress on the structure of the dispersion surfaces (see Fig.~\ref{fig:dispersion_surface_p_0},~\subref*{fig:dispersion_surface_p_65}, and~\subref*{fig:dispersion_surface_p_190}) and consequently to tune the prestress parameter $p$.

\section{Tunable transmission properties of prestressed interfaces}
\label{sec:prestressed_interfaces}
Two forcing sources have been considered to demonstrate the effects related to the presence of an interface separating elastic beams unloaded from beams pre-loaded with an axial force. 

The first dynamic excitation is a plane wave generated and propagated in the grid using the following technique. 
In a first step, by means of Eq.~\eqref{eq:system} a single Floquet-Bloch wave is calculated for an infinite square grid of beams not subject to prestress.
In a second step, the just calculated displacements are applied on a finite portion of the boundary of a square region, which contains layers of prestressed elastic beams and is enclosed within PML boundary conditions (the damping parameter is tuned to prevent reflection at the boundary).
Propagation in the latter square region is analyzed via finite elements using COMSOL Multiphysics$^\circledR$ in the frequency response mode.
The rotational inertia term of the Rayleigh model is implemented by modifying the moment equation of the standard Euler-Bernoulli elements~\cite{piccolroaz_2017a}.
The second dynamic excitation is a concentrated time-harmonic moment (of out-of-plane axis) applied to a junction of the beam network (where the prestress is absent), adjacent to the boundary of the prestressed structured interface. 

In the following applications, the frequency levels have been selected to provide a slowness contour of the ambient lattice (not prestressed) characterized by almost perfectly straight edges, in order to favor a strongly localized forced response~\cite{ruzzene_2003,bordiga_2018}.

\subsection{Total reflection}
\label{sec:total_reflection}
\begin{figure}[htb]
\subfloat{
    \label{fig:reflection_prestress_wave}
    \includegraphics[width=0.48\linewidth]{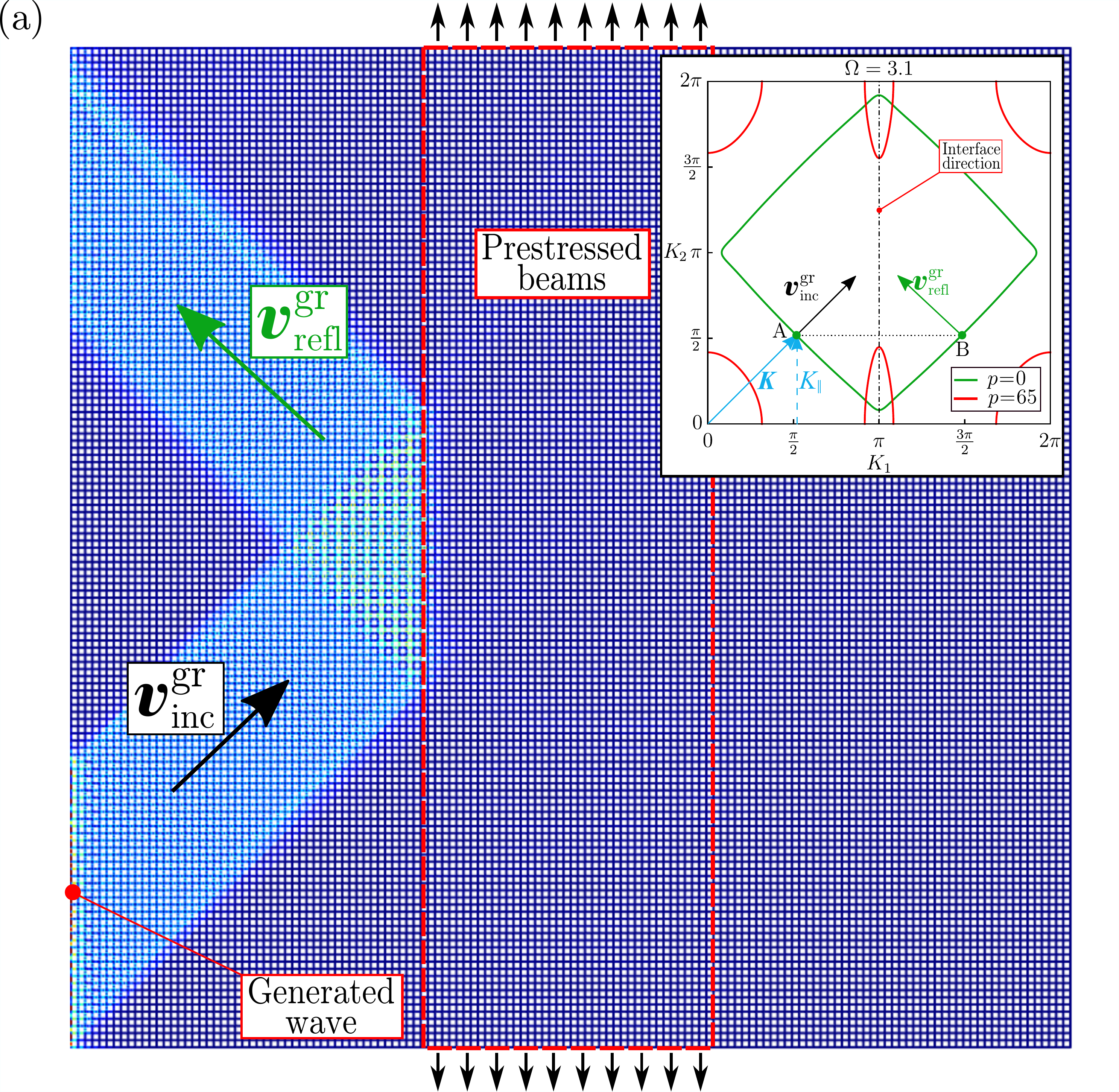}
}
\subfloat{
    \label{fig:reflection_prestress_moment}
    \includegraphics[width=0.48\linewidth]{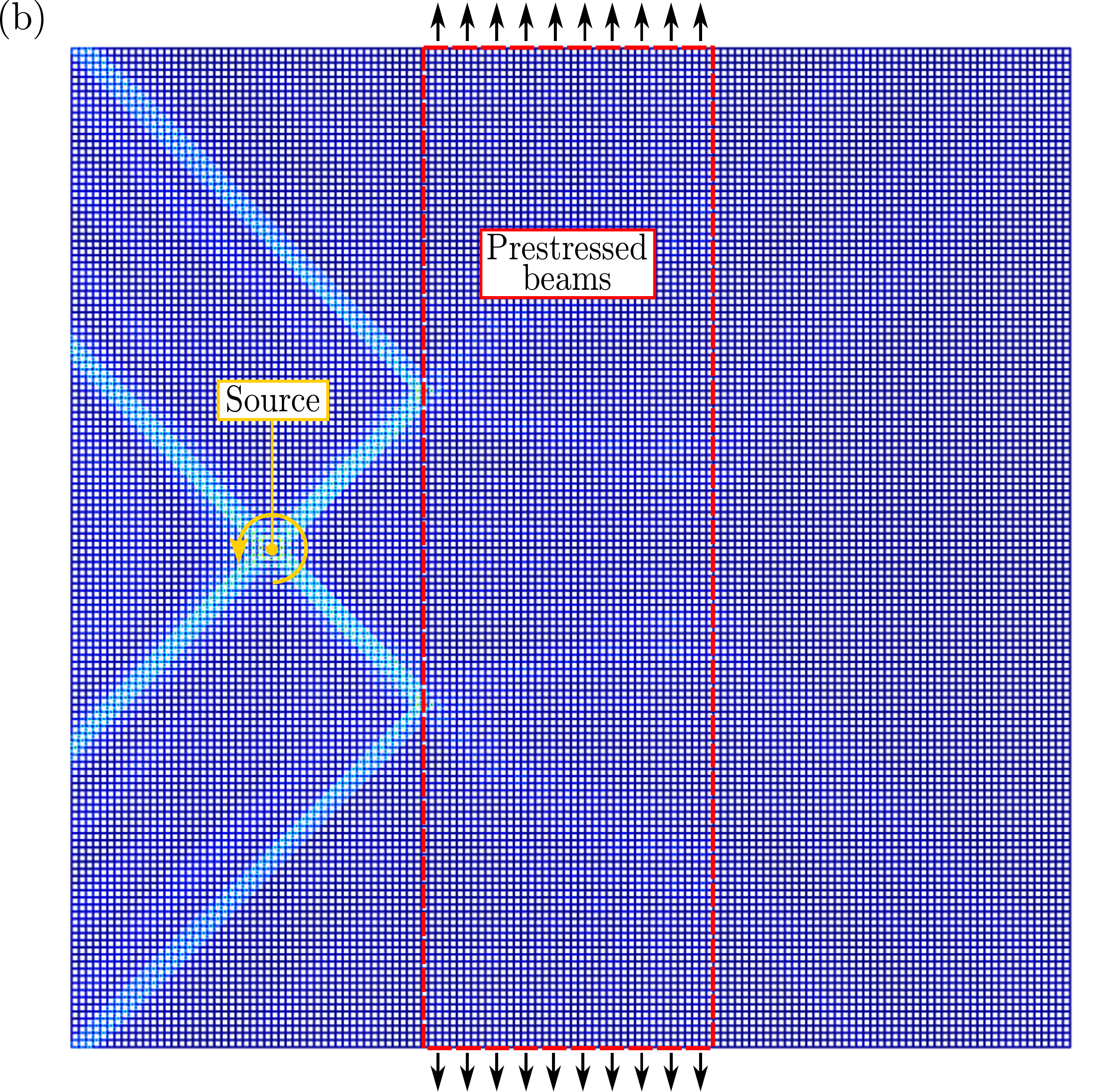}
}
\caption[]{\label{fig:reflection_prestress}Total reflection on a layer of prestressed ($p=65$) elastic rods, at a frequency $\Omega=3.10$ of: \subref{fig:reflection_prestress_wave} a plane wave incident at $45^\circ$ and~\subref{fig:reflection_prestress_moment} a channeled wave pattern generated by a pulsating concentrated moment. 
The black and green arrows in part~\subref{fig:reflection_prestress_wave} denote the group velocities $\bv^\textrm{gr}_\textrm{inc}$ and $\bv^\textrm{gr}_\textrm{refl}$, respectively, and the inset shows the slowness contours for the lattice without (green) and with (red) prestress, respectively. 
}
\end{figure}

A total reflection is shown in Fig.~\ref{fig:reflection_prestress_wave} of a plane wave (inclined at $45^\circ$ and propagating at the frequency $\Omega=3.10$) against an interface with prestressed  vertical elements subject to a tensile load of $p=65$.
This value of prestress (and those assumed in the following) is very high, so that in a practical implementation of the concept presented in this paper, a nonlinear material with a tangent stiffness modulus strongly decreasing with strain has to be used (strictly speaking, only the tangent modulus at the prescribed prestress level enters in the formulation).
% 
%%%%%%%%%%%%%%%%%%%%%%%%%%%%%%%%%%%%%%%%%%%%%%%
%%%%%%%%%%%%%%%%%%%%%%%%%%%%%%%%%%%%%%%%%%%%%%%

The black and green arrows denote the group velocity of the incident wave $\bv^\textrm{gr}_\textrm{inc}$ and of the reflected wave $\bv^\textrm{gr}_\textrm{refl}$, respectively, while the slowness contour of the lattice is shown in the inset without (marked green) and with (marked red) prestress.
With reference to the inset, the propagation direction of the incident plane wave, with fronts perpendicular to vector $\bK$, is defined by the gradient of the dispersion relation at the point $A$, i.e. the group velocity of the incident wave $\bv^\textrm{gr}_\textrm{inc}$ (black arrow) in the lattice without prestress. 
Using the conservation of the component of $\bK$ parallel to the interface ($K_\parallel=K_2$), the gradient at point $B$ determines the group velocity of the reflected wave $\bv^\textrm{gr}_\textrm{refl}$ (green arrow). 
The directions of $\bv^\textrm{gr}_\textrm{inc}$ and $\bv^\textrm{gr}_\textrm{refl}$ highlight the total reflection, also marked by the fact that the slowness contour of the interface (red) is not intersected by the projection of the vector $\bK$ along the direction of the interface.

Channeling of the signal generated by a pulsating concentrated moment (of out-of-plane axis) is shown in Fig.~\ref{fig:reflection_prestress_moment}. 
The source, vibrating at $\Omega=3.10$, is applied near the same interface used for Fig.~\ref{fig:reflection_prestress_wave} and defining a prestressed layer, so that total reflection is again observed, but now obtained for the wide Bloch spectrum generated by the pulsating moment.

\subsection{Negative refraction and flat lens}
\label{sec:negative_refraction_flat_lens}
\begin{figure}[htb]
\subfloat{
    \label{fig:negative_refraction_prestress_wave}
    \includegraphics[width=0.48\linewidth]{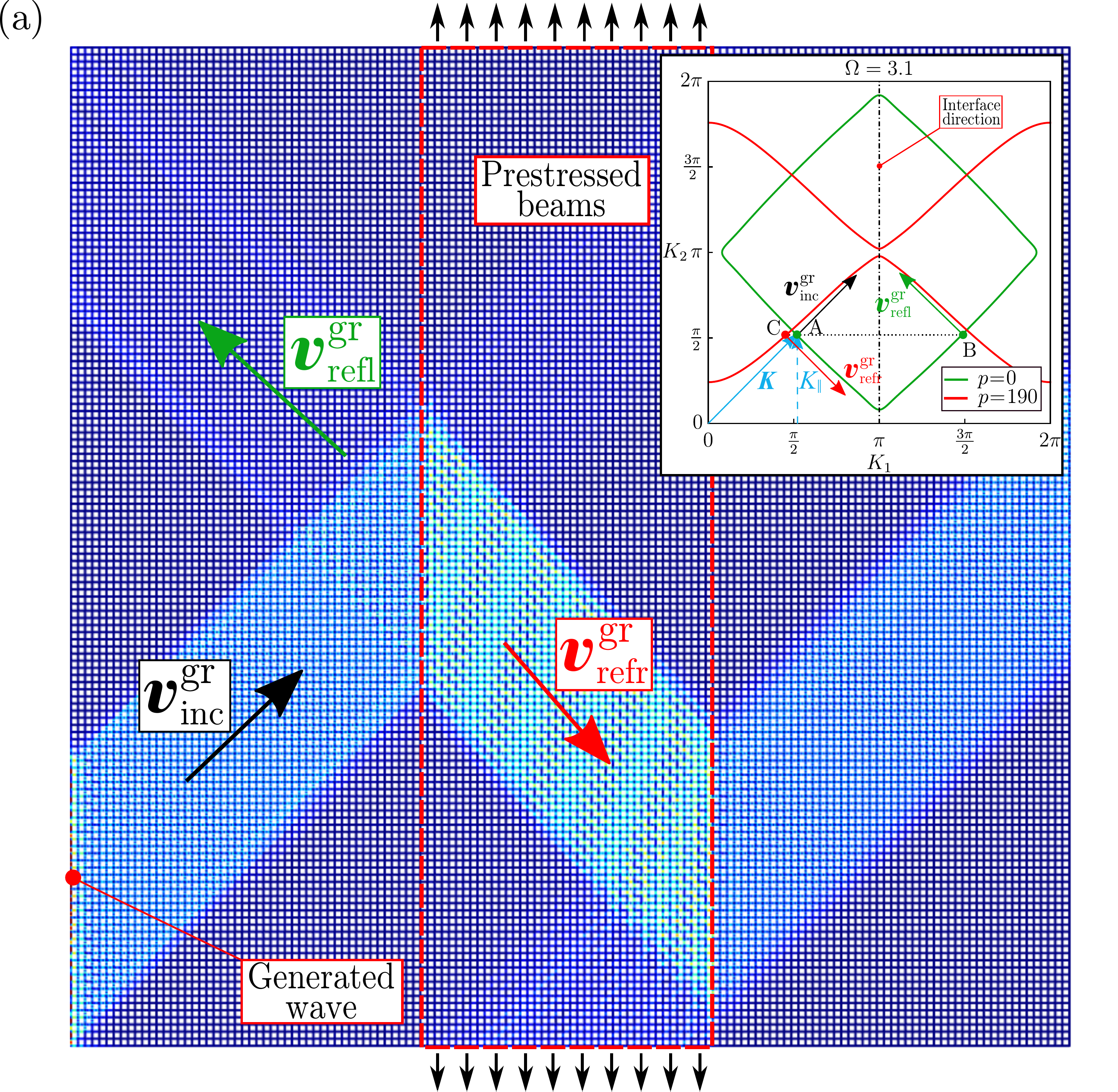}
}
\subfloat{
    \label{fig:negative_refraction_prestress_moment}
    \includegraphics[width=0.48\linewidth]{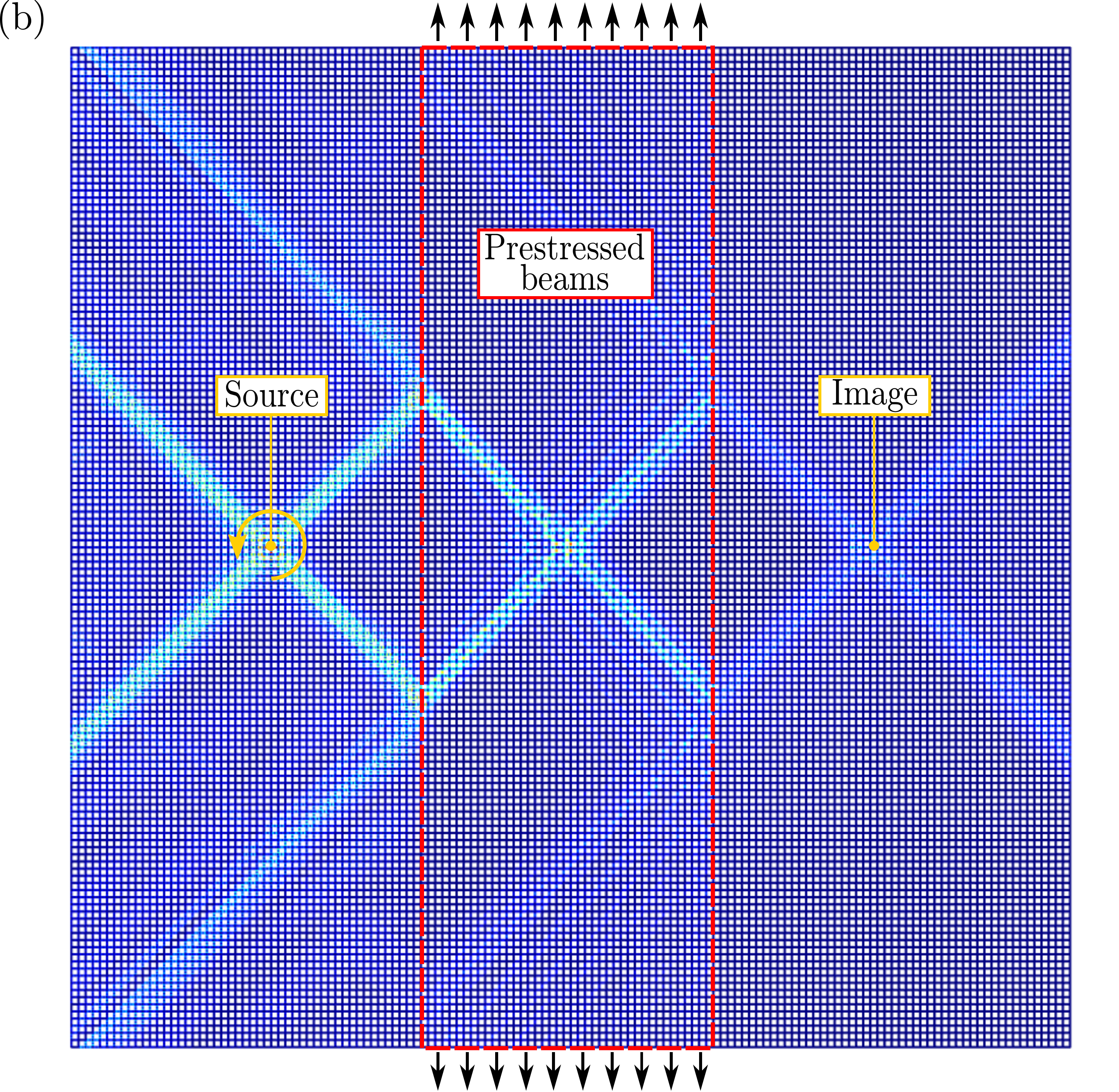}
}
\caption[]{\label{fig:negative_refraction_prestress}Negative refraction inside a layer of highly prestressed, $p=190$, beams~\subref{fig:negative_refraction_prestress_wave}, producing a flat lens~\subref{fig:negative_refraction_prestress_moment}.
In part~\subref{fig:negative_refraction_prestress_wave} a plane wave is incident at $45^\circ$ on the interface at the frequency $\Omega=3.10$ and the black, green and red arrows denote the group velocities $\bv^\textrm{gr}_\textrm{inc}$, $\bv^\textrm{gr}_\textrm{refl}$ and $\bv^\textrm{gr}_\textrm{refr}$, respectively.
The inset shows the slowness contours for the periodic lattice without (green) and with (red) prestress, respectively.
In part~\subref{fig:negative_refraction_prestress_moment} a channeled wave pattern is generated by a pulsating concentrated moment and the corresponding image is reconstructed through a flat lens interface.
}
\end{figure}

Tuning the prestress to $p=190$ in the geometry already analyzed for total reflection, now negative refraction is observed (Fig.~\ref{fig:negative_refraction_prestress_wave}), so that a part of the incident wave continues to be reflected and another part crosses the interface with a strongly negative angle of refraction. 
In particular, the black, the green and the red arrows denote the group velocity of the incident $\bv^\textrm{gr}_\textrm{inc}$, of the reflected $\bv^\textrm{gr}_\textrm{refl}$ and of the refracted $\bv^\textrm{gr}_\textrm{refr}$ waves, respectively.
The inset shows that the green slowness contour remains the same as that of the grid without prestress, while the red contour is now modified by the higher value of prestress $p$. 
The level of prestress is tuned to obtain a significant change of the group velocity direction between the ambient lattice and the prestressed grid.
The gradient at point $B$ determines the group velocity $\bv^\textrm{gr}_\textrm{refl}$ (green arrow), while the gradient at point $C$ the group velocity of the refracted wave $\bv^\textrm{gr}_\textrm{refr}$ (red arrow). 
The strong negative refraction follows from the scalar product $\bv^\textrm{gr}_\textrm{inc}\scalp\bv^\textrm{gr}_\textrm{refr}\approx0$. 
% 
%%%%%%%%%%%%%%%%%%%%%%%%%%%%%%%%%%%%%%%%%%%%%%%
%%%%%%%%%%%%%%%%%%%%%%%%%%%%%%%%%%%%%%%%%%%%%%

Using the negative refraction achieved with the prestressed interface, it is possible to transform the layer of prestressed beams into a flat lens~\cite{veselago_1968,pendry_2000,zhang_2004,brun_2010}, as demonstrated by the wave focusing, forming an image evident in Fig.~\ref{fig:negative_refraction_prestress_moment}, where a channeled wave pattern generated by the pulsating concentrated bending moment is reported. 

It is worth noting that the designed interface is capable of refracting most of the Bloch spectrum activated by the pulsating moment and therefore reflecting only a small part of the incident signal.
Furthermore, as a consequence of the simplicity of the tuning obtained through prestress of some beams, the transmission properties of the interface can be easily changed, so that the response can switch from a pure reflection (Fig.~\ref{fig:reflection_prestress_moment}) to a flat lens effect (Fig.~\ref{fig:negative_refraction_prestress_moment}).
This provides much more flexibility for applications in dynamics than solutions requiring structural modifications of the material~\cite{zhang_2004,hladky-hennion_2008,liu_2011,smith_2012,zhu_2014,matlack_2018}.

\subsection{Trapping and focussing of a signal}
\label{sec:trapping_focussing}
Taking advantage of the interplay between the mechanical properties of the lattice subject or not to prestress, it is possible to introduce a band of prestressed beams inside an homogeneous lattice in such a way to generate complex paths to be followed by a trapped wave generated by a concentrated bending moment applied inside the path.  
% 
%%%%%%%%%%%%%%%%%%%%%%%%%%%%%%%%%%%%%%%%%%%%%%%
\begin{figure}[htb]
\subfloat{
    \label{fig:trapping_prestress}
    \includegraphics[width=0.48\linewidth]{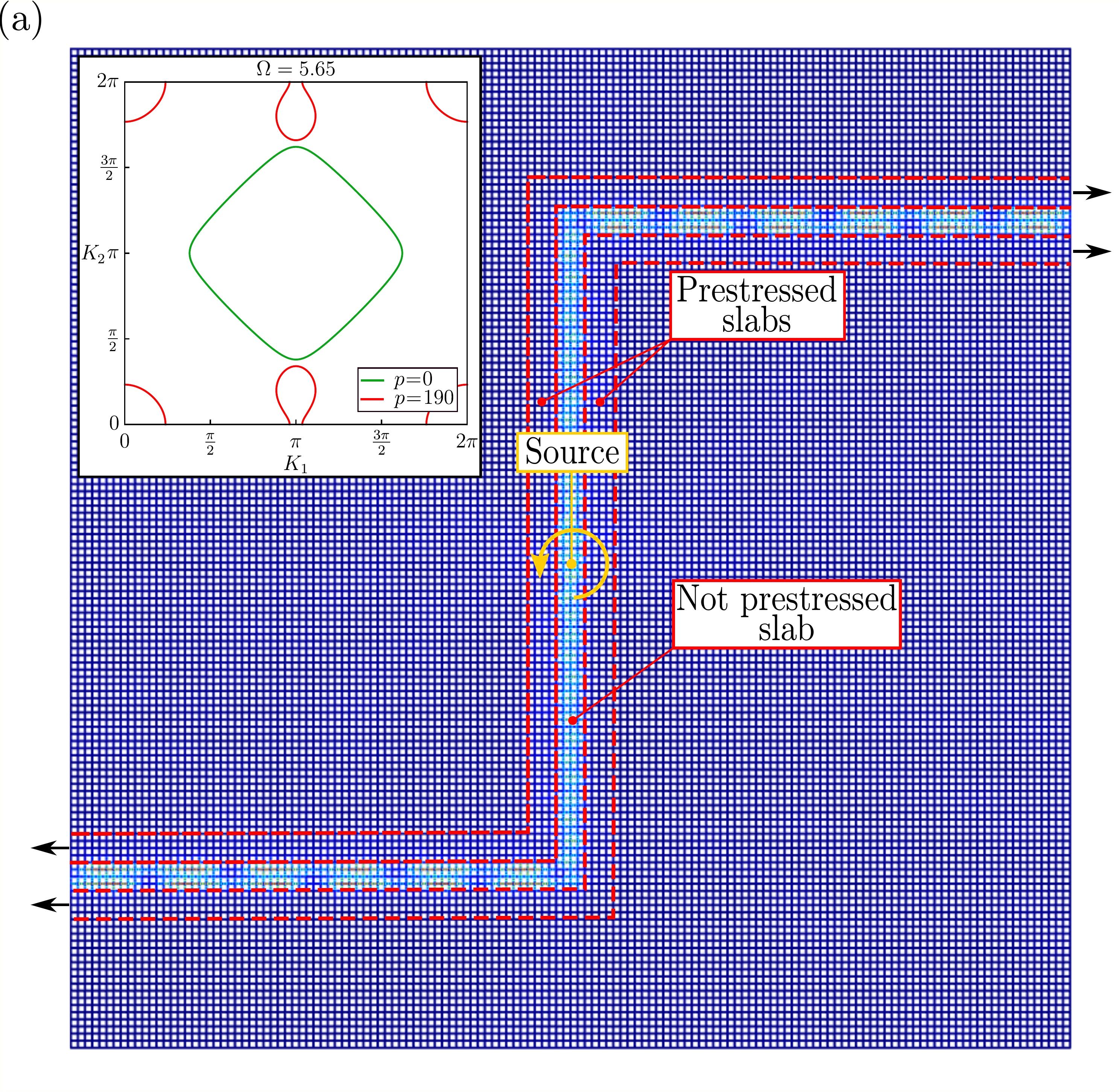}
}
\subfloat{
    \label{fig:focussing_prestress}
    \includegraphics[width=0.48\linewidth]{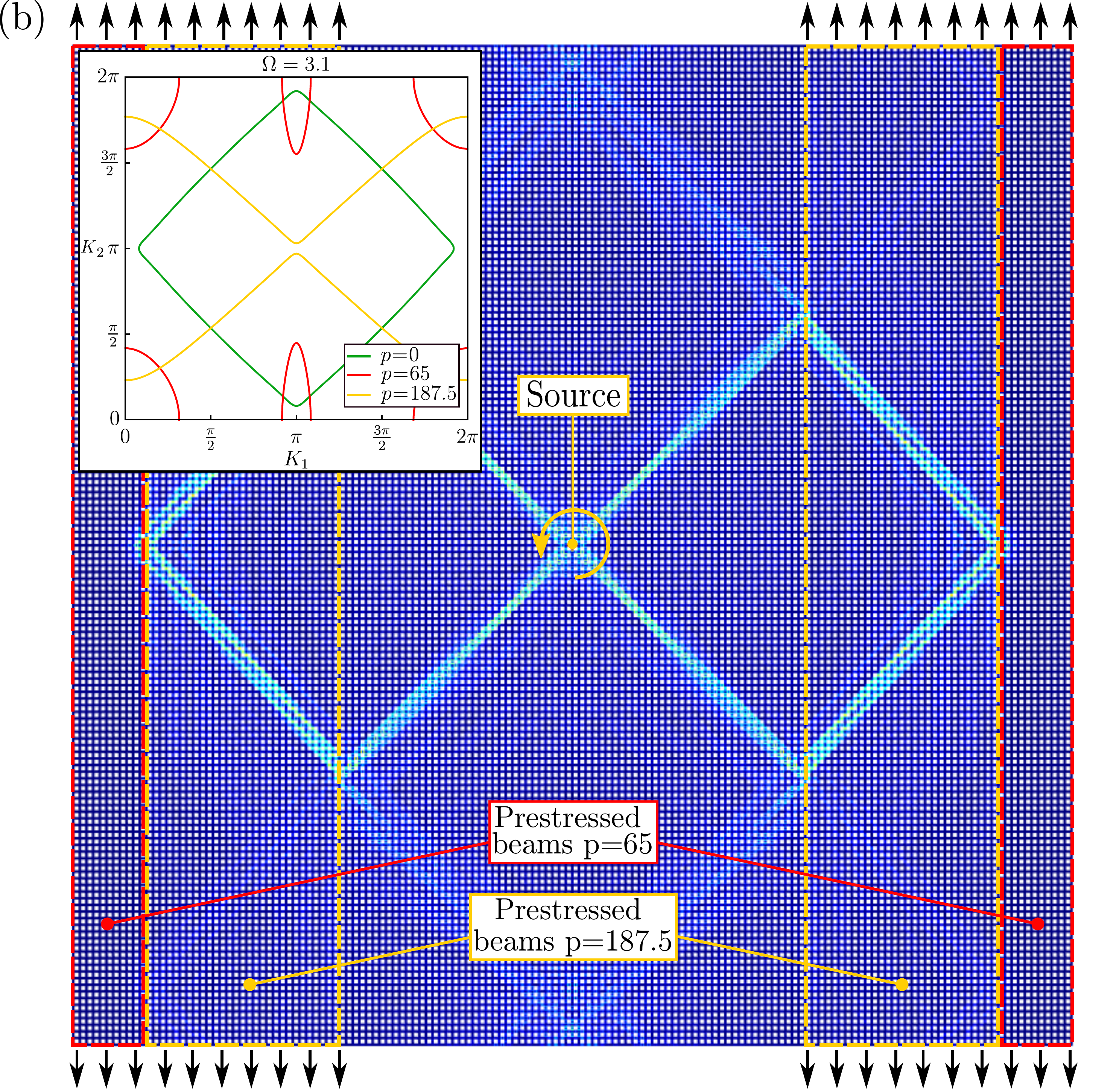}
}
\caption[]{\label{fig:trapping_focussing}\subref{fig:trapping_prestress} Trapping of a wave generated by a concentrated bending moment, pulsating at the frequency $\Omega=5.65$, inside a channel delimited by two narrow layers of prestressed beams ($p=190$) aligned parallel to the desired propagation path. 
\subref{fig:focussing_prestress} The combination of two layers of beams at different level of prestress ($p=65$ and $p=187.5$) generates a complex channeled wave pattern with negative refraction and focussing effect ($\Omega=3.10$).
}
\end{figure}
%%%%%%%%%%%%%%%%%%%%%%%%%%%%%%%%%%%%%%%%%%%%%%%

An `S-shaped' trapped wave is shown in Fig.~\ref{fig:trapping_prestress}, which propagates at frequency $\Omega=5.65$ inside a channel delimited by two narrow layers of prestressed beams ($p=190$), an expedient which realizes a simple method to spatially control the energy transmission along any desired path~\cite{jiao_2018}.
Again, since the effect is prestress-induced, it provides a valuable alternative to other methods of channeling dynamic signals (e.g. leveraging edge-waves in gyroscopic systems~\cite{wang_2015,carta_2017,garau_2018} and topological materials~\cite{xia_2018} or embedding piezoelectric elements in elastic lattices~\cite{celli_2015}).

Finally, it is worth mentioning that layers of beams subject to different prestress levels can be introduced to obtain complicated effects. 
For instance, a complex channeled wave pattern showing negative refraction and focussing effects is shown in Fig.~\ref{fig:focussing_prestress}, as obtained by the combination of two layers of prestressed beams at different values of force ($p=65$ and $p=187.5$).
This double-layer interface is designed to first bend the signal (through the negative refraction occurring at the first prestressed layer $p=187.5$) and then to focus it on the reflective prestressed layer ($p=65$). 
The second interface totally reflects the signal towards the first one, which in turn directs it, with a negative angle of refraction, towards the original source, where  the signal is concentrated exactly where it is generated.
Only a small part of the original signal is lost in the passage through the first interface due to partial reflection. 
As the thickness of these layers can be easily adjusted, the point of focussing can be effectively engineered as a function of distance from the forcing source.

\section{Concluding remark}
\label{sec:conclusion}
In summary, we have demonstrated that prestress represents a simple way to tune the mechanical properties of an elastic grid of (axially and flexurally deformable) beams, so that the response to wave propagation of a totally reflective interface can be changed so to leave the signal refracting through the interface with a negative angle. 
Moreover, the prestress can be used to localize wave propagation into narrow layers inside a material, to mimic edge wave propagation in topological materials, or to trap energy inside thin channels. 
The wave manipulation tool proposed in this letter benefits from the fact that the signal is generated by a moment source and the obtained dynamical properties work correctly for a wide range of wavelengths.

\section*{Supplementary material}
Results similar to those reported in the Letter can also be obtained without prestress, but through a proper use of the slenderness $\lambda$ of the beams. In this case, the tunability of the structure is lost. 
 
\begin{figure}[htb]
\centering
\subfloat[][]{
    \label{fig:SlendernessReflection}
    \includegraphics[width=0.31\linewidth]{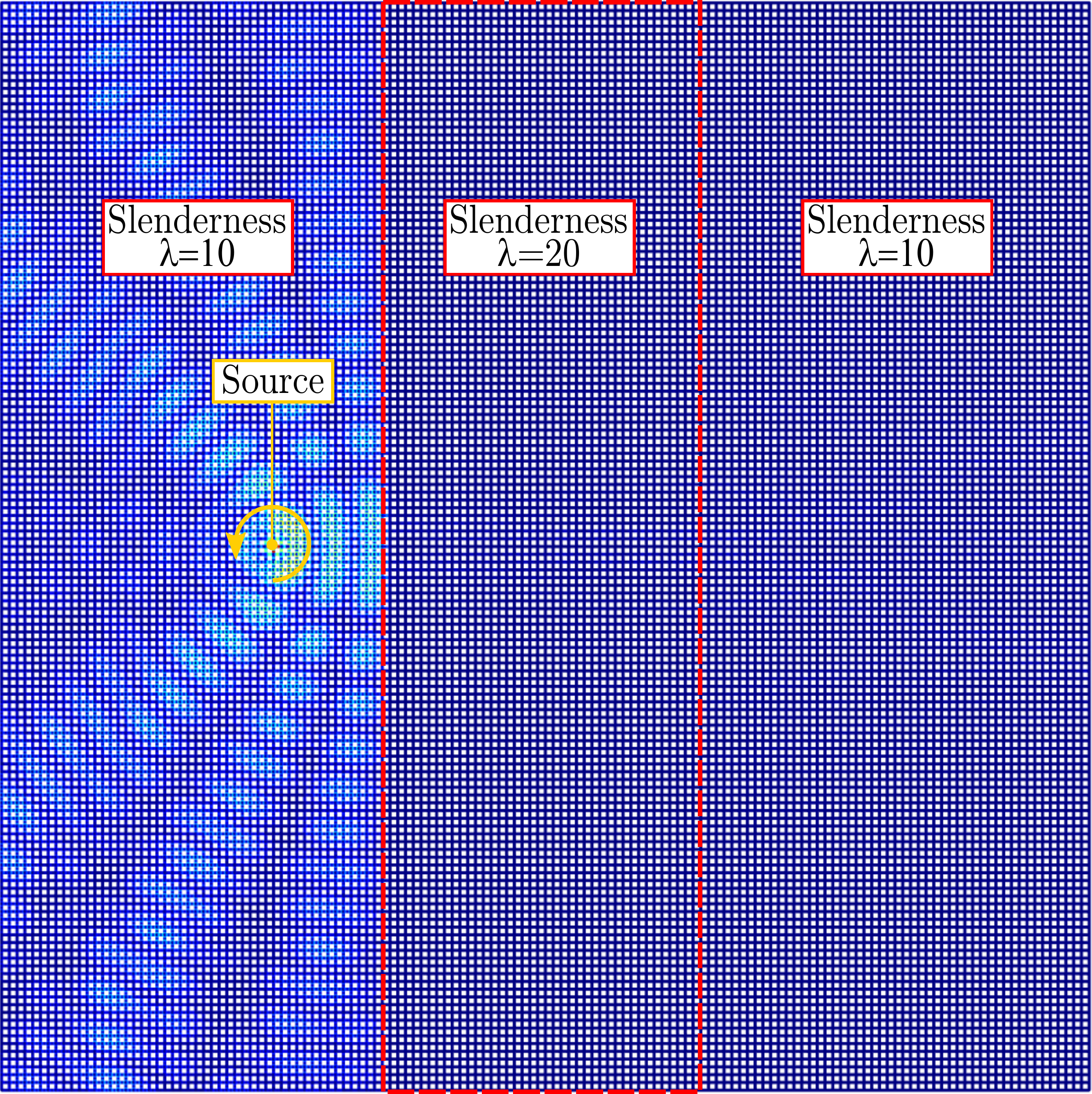}
}\hfill
\subfloat[][]{
    \label{fig:SlendernessTrapping}
    \includegraphics[width=0.31\linewidth]{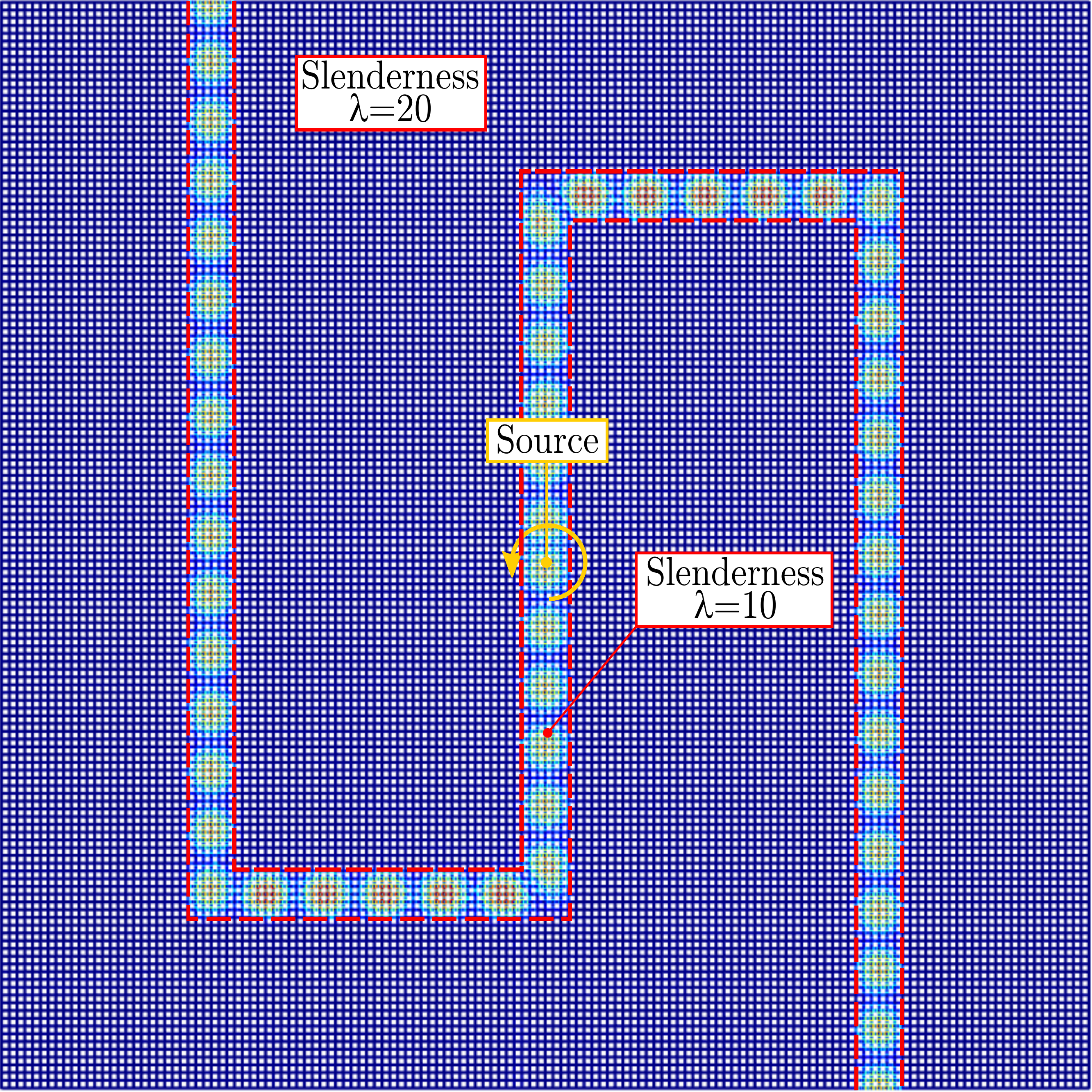}
}\hfill
\subfloat[][]{
    \label{fig:SlendernessRefraction}
    \includegraphics[width=0.31\linewidth]{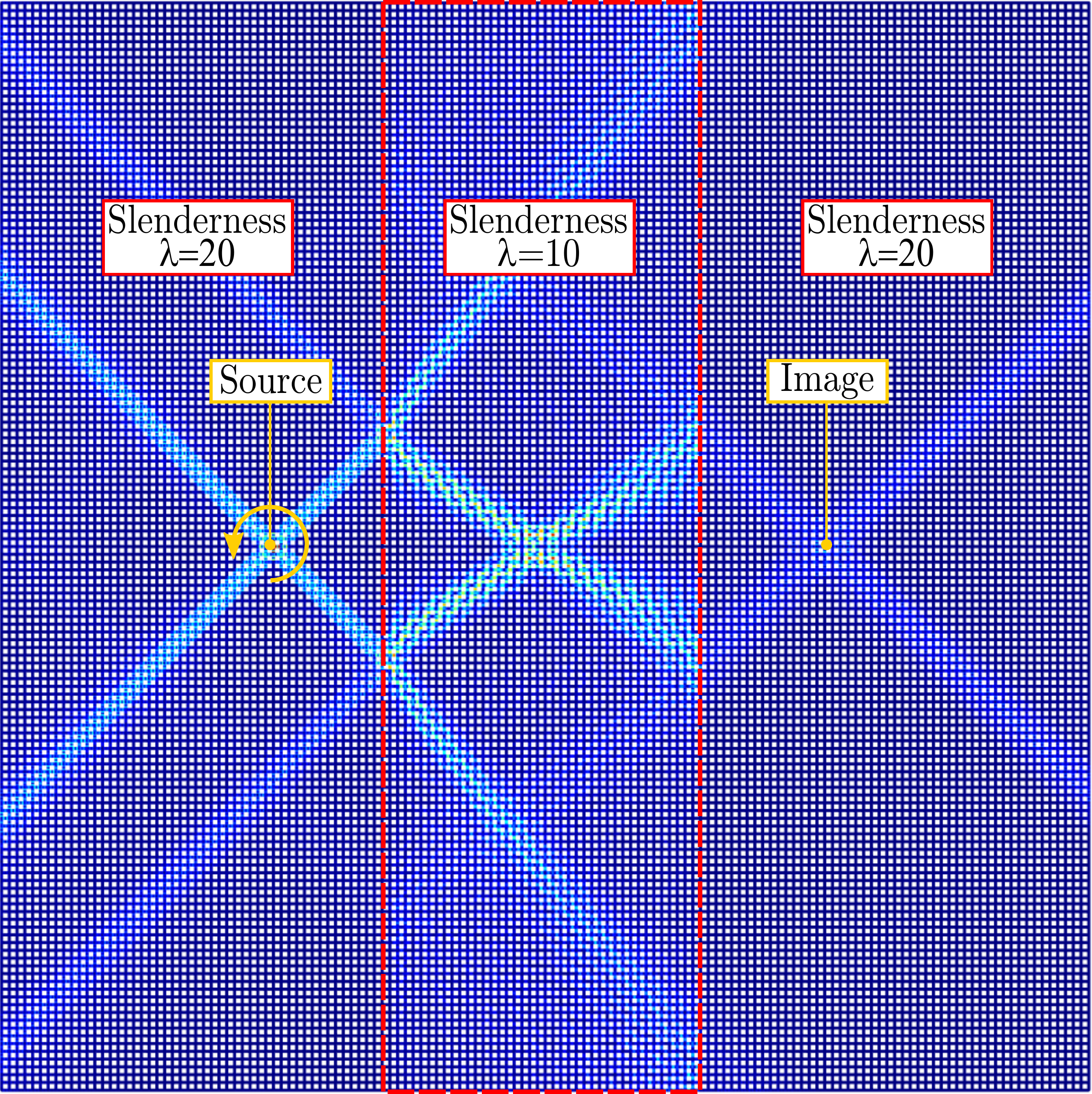}
}\hfill
\caption[]{\label{fig:Slenderness}(a) Total reflection of a signal generated by a moment (which pulsates at the frequency $\Omega=3.43$ and is applied to a lattice with beams of slenderness $\lambda=10$) at a structured interface made of beams with slenderness $\lambda=20$.
(b) Trapping of a wave generated by an applied moment, located inside a narrow layer of beams of slenderness $\lambda=10$ embedded in a lattice with slenderness $\lambda=20$. 
(c) Negative refraction of a wave generated by an applied moment (pulsating at the frequency $\Omega=2.5$), located inside a layer of beams of slenderness $\lambda=10$, embedded in a lattice of beams with $\lambda=20$.
}
\end{figure}

Fig.~\ref{fig:SlendernessReflection} shows how a signal induced by a moment (of out-of-plane axis), pulsating at the frequency $\Omega=3.43$, applied to a lattice with beams of slenderness $\lambda=10$ is totally reflected by a structured interface made of beams with slenderness $\lambda=20$. 
Taking advantage of this total reflection effect, it is possible to channel waves along a path by applying a pulsating moment, inside of a layer of beams with slenderness $\lambda=10$, embedded in a lattice of beams with $\lambda=20$, as highlighted in Fig.~\ref{fig:SlendernessTrapping}. 

By swapping the slenderness values of the ambient lattice and the structured interface of the above example, the negative refraction effect is obtained, at a frequency $\Omega=2.5$, Fig.~\ref{fig:SlendernessRefraction}. 
Note that the dynamic response induced by a pulsating moment displays a strong localization along two preferential directions inclined at $\pm45^\circ$. 
Part of the generated waves is reflected by the interface and part is transmitted. 
It is possible to see how the refracted waves is essentially split into two channels: one with positive and the other with negative angle of refraction. 
The different angles of refraction are due to the different response of the single Bloch-wave components generated at the source.

\vspace{10mm}
{\small
\noindent
\noindent
\textbf{Funding.} G.B., L.C., D.B., gratefully acknowledge financial support from the ERC Advanced Grant `Instabilities and nonlocal multiscale modelling of materials' ERC-2013-ADG-340561-INSTABILITIES.
A.P. thanks financial support from the PRIN 2015 `Multi-scale mechanical models for the design and optimization of micro-structured smart materials and metamaterials' 2015LYYXA8-006.
}

\printbibliography

\end{document}